\documentstyle[12pt,aasms4,flushrt]{article}

\def\gsim{\;\rlap{\lower 2.5pt
 \hbox{$\sim$}}\raise 1.5pt\hbox{$>$}\;}
\def\lsim{\;\rlap{\lower 2.5pt
   \hbox{$\sim$}}\raise 1.5pt\hbox{$<$}\;}
\newcommand\beq{\begin{equation}}
\newcommand\eeq{\end{equation}}

\def\v{\vspace{-0.1in}}

\begin{document}
\Large 
\centerline{\bf The Formation of the First Low-Mass Stars From}
\centerline{\bf Gas With Low Carbon and Oxygen Abundances}
\normalsize 
\author{\bf Volker Bromm and Abraham Loeb}

\noindent
Astronomy Department, Harvard University, 60 Garden Street,
Cambridge, MA 02138, USA

\bigskip
\centerline{Appeared in {\it Nature}, {\bf 425}, 812--814 (2003)}
\medskip

\vskip 0.2in 
\hrule 
\vskip 0.2in 
\noindent
{\bf The first stars in the Universe are predicted to have been much more
massive than the Sun\cite{abn02}$^-$\cite{nu02}. Gravitational condensation
accompanied by cooling of the primordial gas due to molecular hydrogen,
yields a minimum fragmentation scale of a few hundred solar
masses. Numerical simulations indicate that once a gas clump acquires this
mass, it undergoes a slow, quasi-hydrostatic contraction without further
fragmentation\cite{abn02}$^,$\cite{bcl02}.  
Here we show that as soon as the primordial gas -- left over from the Big
Bang -- is enriched by supernovae to a carbon or oxygen abundance as small
as $\sim 0.01$--$0.1\%$ of that found in the Sun, cooling by singly-ionized
carbon or neutral oxygen can lead to the formation of low-mass stars.  This
mechanism naturally accommodates the discovery\cite{HEStar} of solar mass
stars with unusually low ($10^{-5.3}$ of the solar value) iron abundance
but with a high ($10^{-1.3}$ solar) carbon abundance.  The minimum stellar
mass at early epochs is partially regulated by the temperature of the
cosmic microwave background.  The derived critical abundances can be used
to identify those metal-poor stars in our Milky Way galaxy with elemental
patterns imprinted by the first supernovae.}

The microphysics of cooling due to molecular hydrogen (H$_{2}$) within the
first gas clouds to condense in the Universe, introduces a characteristic
temperature ($T_c \sim 100$--$200$~K) and hydrogen density ($n_c \sim
10^{4}$~cm$^{-3}$) into the formation process of the first
stars\cite{abn02}$^,$\cite{bcl02}. The temperature floor is set by the
energy separation of the lowest-lying rotational levels of H$_{2}$, and the
characteristic density leads to the thermalization of these levels, at
which point cooling becomes less efficient.  The minimum mass scale for
fragmentation is approximately given by the Jeans mass, $M_{J}\simeq 400
M_{\odot}(T/200{\rm\, K})^{3/2}(n/10^{4}{\rm cm}^{-3})^{-1/2}$.  The final
mass of a star is determined by the accretion process onto the nascent
protostellar core, resulting in $M_{\ast}\simeq \alpha M_{J}$ with $\alpha
\la 0.5$ \cite{mkt02}$^,$\cite{cb03}.  In contrast to the formation mode of
massive stars (population~III) at high redshifts, fragmentation is observed
to favor stars below a solar mass (population I and II) in the present-day
Universe.  The transition between these fundamental modes is expected to be
mainly driven by the progressive enrichment of the cosmic gas with heavy
elements (or `metals'), which enable the gas to cool to lower temperatures.
The concept of a `critical metallicity', $Z_{\rm crit}$, has been
used\cite{om00}$^-$\cite{sch03} to characterize the transition between
population~III and population~II formation modes (where $Z$ denotes the
mass fraction contributed by all heavy elements). Previous
studies\cite{om00}$^,$\cite{bfcl01}$^,$\cite{sch03} have only constrained
this important parameter to within a few orders of magnitude, $Z_{\rm
crit}\sim 10^{-6}-10^{-3} Z_{\odot}$, under the implicit assumption of
solar relative abundances of metals. This assumption is likely to be
violated by the metal yields of the first supernovae (SNe) at
high-redshifts, for which strong deviations from solar abundance ratios are
predicted\cite{hw02}$^-$\cite{un03}.  The cooling rate of the metals
depends on their ionization state, which in turn is controlled by the
ionizing backgrounds (UV and X-ray photons or cosmic rays) that were not
fully considered in earlier studies.

Here we show that the transition between the above star formation modes is
driven primarily by fine-structure line cooling of singly-ionized carbon or
neutral atomic oxygen.  We refine earlier estimates of $Z_{\rm crit}$ which
did not explicitly distinguish between different coolants, by introducing
separate critical abundances for carbon and oxygen, [C/H]$_{\rm crit}$ and
[O/H]$_{\rm crit}$, respectively, where [A/H]= $\log_{10}(N_{\rm A}/N_{\rm
H})-\log_{10}(N_{\rm A}/N_{\rm H})_{\odot}$, and a subscript `$\odot$'
denotes solar values. {\it Why are C and O identified as the key species
responsible for the global shift in the star formation mode?}  We find that
under the temperature and density conditions that characterize
population~III star formation, the most important coolants are O\,I and
C\,II whose fine-structure lines dominate over all other metal
transitions\cite{hmk89}. Cooling due to molecules becomes important only at
lower temperatures, and cooling due to dust grains only at higher
densities\cite{hmk89}$^,$\cite{om00}$^,$\cite{sch03}.

The conditions that terminate the predominance of massive, population~III
stars, have dramatic implications for the reionization history of the
Universe.  Massive stars are much more efficient at producing ionizing
radiation than low mass stars, due to their high surface temperature ($\sim
10^5$\,K).  In particular, metal-free stars with a mass $\ga 100M_\odot$
produce $\sim 5\times 10^4$ hydrogen ionizing photons per baryon
incorporated into them, a yield higher by an order of magnitude than
obtained for a present-day mass function of
stars\cite{ts00}$^,$\cite{bkl01}. Massive stars can therefore reionize the
universe early\cite{cen03}$^-$\cite{wl03} at $z\sim 17$, as required by the
recent detection\cite{kog03} of large scale polarization anisotropies in
the cosmic microwave background (CMB).  A strong UV flux just below the
Lyman limit ($h\nu < 13.6$\,eV) is predicted\cite{bl03} to be emitted by
the same stars that are responsible for the reionization of the IGM. This
soft UV radiation can penetrate the neutral IGM and ionize any trace amount
of neutral carbon due to its low first-ionization potential of
11.26\,eV. Since carbon is highly underabundant, the UV background can
ionize carbon throughout the universe well before hydrogen reionization.
Oxygen, on the other hand, will be predominantly neutral prior to
reionization since its ionization potential is 13.62\,eV.  Cooling is
mediated by excitations due to collisions of the respective metal with free
electrons or hydrogen atoms. At the low fractional abundances of electrons,
$x_{e}\la 10^{-4}$, expected in the neutral (rapidly recombining) gas at
$z\sim 15$, collisions with hydrogen atoms dominate. Even if the gas had
initially been ionized to a much higher degree due to the impact of the
supernova ejecta, it would have recombined to the level stated above within
its free-fall time [$t_{\rm ff}\sim 5\times 10^{5}
(n/10^{4}$\,cm$^{-3})^{-1/2}$\,yr].  This renders our analysis independent
of the very uncertain hydrogen-ionizing backgrounds at high redshifts
(cosmic rays or soft X-rays).

We now derive the critical C and O abundances. Our starting point is the
characteristic state reached in primordial gas that cools only through
molecular hydrogen. For the gas to fragment further (instead of following a
slow contraction to a single massive star\cite{abn02}$^,$\cite{bcl02}),
additional cooling due to C\,II or O\,I is required. Fragmentation requires
that the radiative cooling rate be higher than the free-fall compressional
heating rate. We find the critical metal abundances by equating the two
rates: $\Lambda_{\rm C\,II, O\,I}(n,T)\simeq 1.5 n k_{\rm B} T/t_{\rm ff}$,
where $k_{\rm B}$ is Boltzmann's constant.  In evaluating this threshold
condition, we solve the respective rate equations for C\,II (a two-level
system) and O\,I (a three-level system), including all possible radiative
and collisional transitions\cite{hmk89}. We assume that the emission of the
fine-structure lines proceeds under optically thin conditions, as
appropriate for the low atomic abundances considered here\cite{hmk89}. We
find the critical C and O abundances by setting $n\sim n_c$ and $T \sim
T_c$, and estimate the uncertainties by considering a plausible range near
these values ($100{\rm \,K}\la T\la 200{\rm \,K}$).  

In Fig.~1, we show the threshold C and O abundances that enable continued
fragmentation for various temperature and density values. It is convenient
to plot these quantities as functions of the Jeans mass ($\propto
T^{3/2}n^{-1/2}$).  We find critical abundance values of [C/H]$_{\rm
crit}\simeq -3.5 \pm 0.1$ and [O/H]$_{\rm crit}\simeq -3.05 \pm 0.2$.
Strictly speaking, these threshold levels are the required abundances in
the gas phase. If a fraction of these elements were depleted onto dust
grains, then the total C and O elemental abundances would have to be
correspondingly higher. Considering the large current uncertainties in
modelling the properties and formation channels of dust at low metal
abundances\cite{sch03}, we assume the limiting case of zero C and O
depletion onto dust when comparing our theoretical predictions with
observed stellar abundances.

We stress that even if sufficient C or O atoms are present to further cool
the gas, there will be a minimum attainable
temperature\cite{lar98}$^,$\cite{cb03} that is set by the interaction of
the atoms with the thermal CMB: $T_{\rm CMB}=2.7{\rm \,K}(1+z)$. At
$z\simeq 15$, this results in a minimum fragment mass of $M_{\ast}\sim 20
M_{\odot}(n_f/ 10^{4}{\rm \,cm}^{-3})^{-1/2}$, where $n_f> 10^{4}{\rm
\,cm}^{-3}$ is the density at which opacity prevents further
fragmentation\cite{rees76}.  It is possible that the transition from the
high-mass to the low-mass star formation mode was modulated by the CMB
temperature and was therefore gradual\cite{cb03}, involving
intermediate-mass (`population~II.5') stars\cite{mbh03} at intermediate
redshifts. This transitional population can give rise to the faint SNe that
have been proposed to explain the observed abundance patterns in metal-poor
stars\cite{un02}$^,$\cite{un03}. These solar-mass stars themselves should
have formed out of metal-poor gas at relatively late times when the CMB
temperature was sufficiently low.

In relating our results to theoretical nucleosynthetic yields, one has to
address the degree of mixing, or dilution, that the SN ejecta
experience. SNe at high redshifts occur in dark matter halos with shallow
potential wells, so that the resulting blast wave is predicted\cite{byh03}
to escape into the expanding IGM.  The dilution mass can then be estimated
as $M_{\rm dil}\simeq 2 E_{\rm SN}/ v^2$, where $E_{\rm SN}$ is the
explosion kinetic energy, and $v=H(z)r$ is the Hubble velocity at a
distance $r$. Using standard cosmological parameters and ignoring radiative
losses\cite{byh03}, we get $M_{\rm dil}\simeq 10^{7} M_{\odot} (E_{\rm
SN}/10^{53}{\rm\, ergs})^{3/5}[(1+z)/16]^{-3/5}$.  In the expected mass
range of the pair-instability supernovae\cite{hw02} (PISNe) and the
corresponding explosion energies $E_{\rm SN}\simeq 10^{51} -10^{53}$~ergs,
one finds dilution masses that are more than two orders of magnitude larger
than typical values for type~II SNe in nearby galaxies\cite{qw02}. However,
depending on the exact values of the ejected mass of C and O and the
respective explosion energies, a single SN event could already lead to
[C/H] and [O/H] ratios that `overshoot' the critical abundances. For
example, the abundances resulting from the explosion and subsequent mixing
of a PISN at the low-mass end\cite{hw02}, are [C/H]$\simeq -2.9 $ and
[O/H]$\simeq -2.4$, respectively, roughly $0.5$~dex above the critical
levels. Nevertheless, a single PISN at the high-mass end would produce final,
mixed abundances still below critical. Since the dilution mass scales as
the number of simultaneous SNe to the ${3/5}$ power while the ejecta mass
scales linearly with this number, multiple simultaneous SNe within the same
galaxy would be less effective at diluting their metal abundances into the
expanding IGM.  But further dilution may occur during the hierarchical
assembly of galaxies, as infall may add metal-free gas from the IGM to the
enriched interstellar medium.

In Fig.~2 we compare our theoretical thresholds to the observed C and O
abundances in metal-poor dwarf\cite{aker03} and giant\cite{cay03} stars in
the halo of our Galaxy.  As can be seen, all data points lie above the
critical O abundance but a few cases lie below the critical C threshold.
All of these low mass stars are consistent with our model since the
corresponding O abundances lie above the predicted threshold.  The
sub-critical [C/H] abundances could have either originated in the
progenitor cloud or from the mixing of CNO-processed material (with carbon
converted into nitrogen) into the stellar atmosphere during the red giant
phase. To guard against this a~posteriori processing of C, dwarf stars
should be preferred as they provide a more reliable record of the
primordial abundance pattern that was present at birth.  However, since
dwarfs are fainter, another strategy would be to focus on giants which show
evidence for weak mixing (e.g. in terms of high $^{12}$C/$^{13}$C isotope
ratio or low $^{14}$N abundance).  The current data on dwarf stars (filled
symbols in Fig. 2) does not yet reach C and O abundances that are
sufficiently low to probe the theoretical cooling thresholds.  Note also
that the extremely iron-poor star HE0107-5240 has C and O abundances that
both lie above the respective critical levels (Bessell, M., et al., in
preparation). The formation of this low mass star ($\sim 0.8 M_{\odot}$) is
therefore consistent with the theoretical framework considered
here\cite{sch03}$^,$\cite{un03}.

Our model suggests a powerful new probe of the first SNe to have exploded
in the Universe. By selecting an unbiased new sample of Galactic halo 
stars with C or O abundances close to the theoretical critical values, it
should be possible to sample individual SN events originating in the
earliest epochs of cosmic star formation.  The value of [C/Fe] would be
indicative of the particular SN type.  For example, the very high value of
[C/Fe]$\simeq +4$ observed\cite{HEStar} in HE0107-5240, can be explained by
a core collapse SN which leaves behind a black hole after fallback of most
of the outer stellar envelope\cite{un03}. Modern surveys select metal-poor
stars based on their weak or absent Ca II K-lines. Our model suggests that a
better strategy to select truly `second generation' stars is to look for
stars with the lowest abundances of carbon and oxygen.

\vskip 0.2in
\noindent
{\bf Acknowledgments} The authors thank Rennan Barkana and particularly Tim
Beers for helpful discussions, and are grateful to Tim Beers and Max
Pettini for making their data available prior to publication.  This work
was supported in part by NSF, NASA, and the Guggenheim foundation (for
A.L.).

\vskip 0.2in
\noindent 
{\bf Correspondence} and requests for materials should be addressed to
V.B. \\
(email: vbromm@cfa.harvard.edu) or A.L. (email: loeb@cfa.harvard.edu).

\vskip 1in

\begin{figure*}[hptb]
\caption{Required carbon and oxygen abundances (relative to solar values)
for cooling a clump of metal-poor gas faster than its free-fall
(compressional heating) time. The minimum abundances are shown as functions
of the initial Jeans mass of the clump (with higher abundances allowing the
clump to cool and further fragment). The upper panel shows the case of
singly ionized carbon, and the lower panel refers to neutral oxygen. The
curves are labeled by the hydrogen number density in units of cm$^{-3}$,
with the solid curve marking the characteristic density in pre-stellar
clumps of metal-poor gas, $n_c\sim 10^{4}$~cm$^{-3}$. Heavy filled dots
indicate particular temperature values. The critical abundance is
delineated by the horizontal dashed lines, with a heavy line indicating the
central value and the adjacent light lines bracketing the uncertainty
within the characteristic temperature range, $T_c\sim 100$--$200$~K, and
density, $\sim 10^{4}$~cm$^{-3}$, of pre-stellar clumps. The figure
indicates the minimum metal abundance at which fragmentation into low-mass
stars starts, but it does not provide information about the minimum mass of
the fragments at the end of the cooling process.}
\end{figure*}

\begin{figure*}[hptb]
\plotone{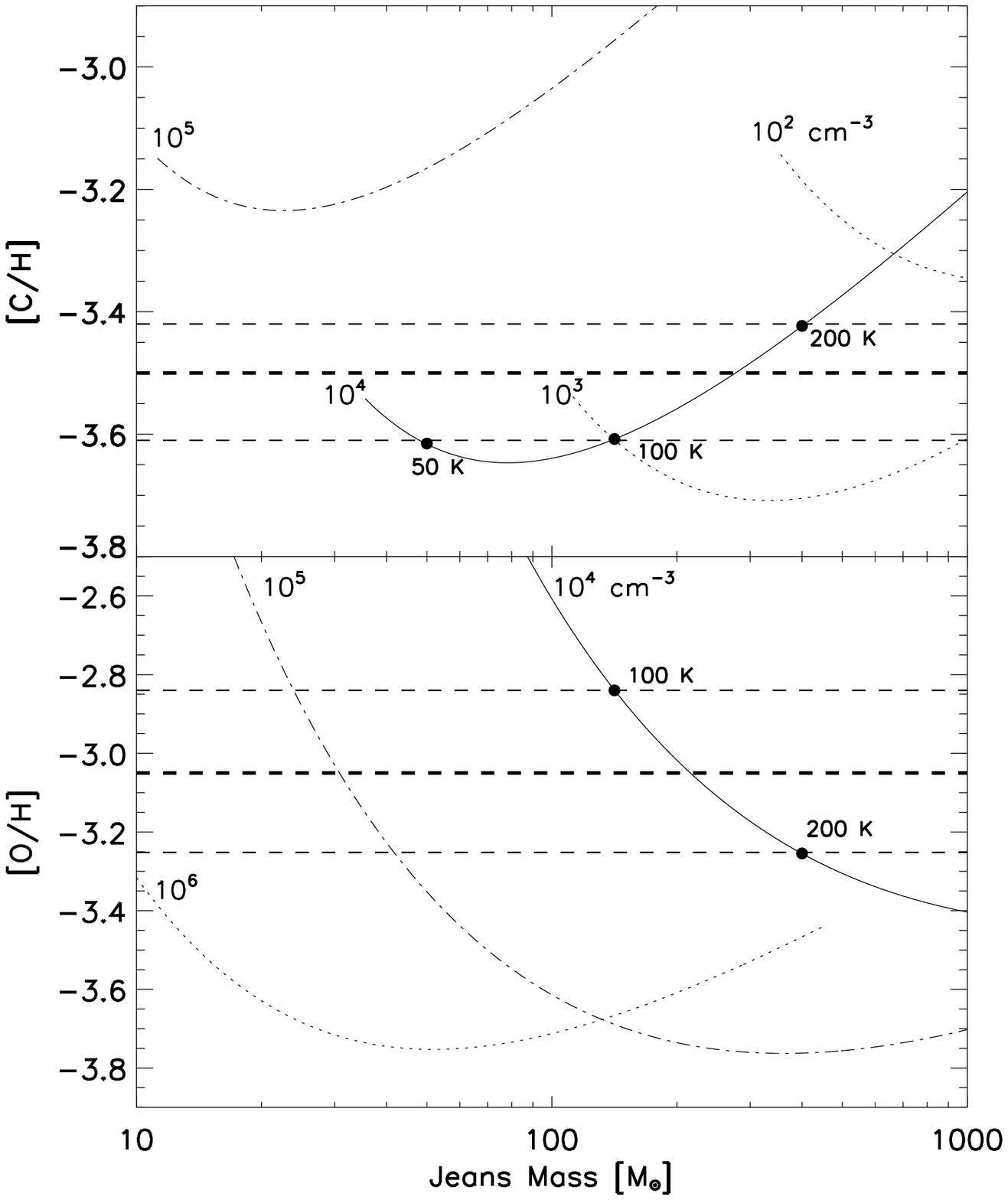}
\end{figure*}

\begin{figure*}[hptb]
\caption{Observed abundances in low-metallicity Galactic halo stars.  For
both carbon (upper panel) and oxygen (lower panel), filled symbols
correspond to samples of dwarf and subgiant stars\cite{aker03}, and open
squares to a sample of giant stars\cite{cay03}.  Both data sets were
obtained at a high signal-to-noise ratio using the UVES spectrograph on the
VLT.  The dashed lines indicate the predicted level of critical carbon and
oxygen abundances from Fig.~1. According to our theoretical model, all
low-mass dwarf stars ought to lie above the critical threshold level of at
least one of the two elements. Indeed all stars in the existing data sets
satisfy this condition. We highlight the location of the extremely
iron-poor giant star HE0107-5240 (marked by {\bf x}), whose carbon and
oxygen abundances are both above the critical levels\cite{HEStar}.}
\end{figure*}

\begin{figure*}[hptb]
\plotone{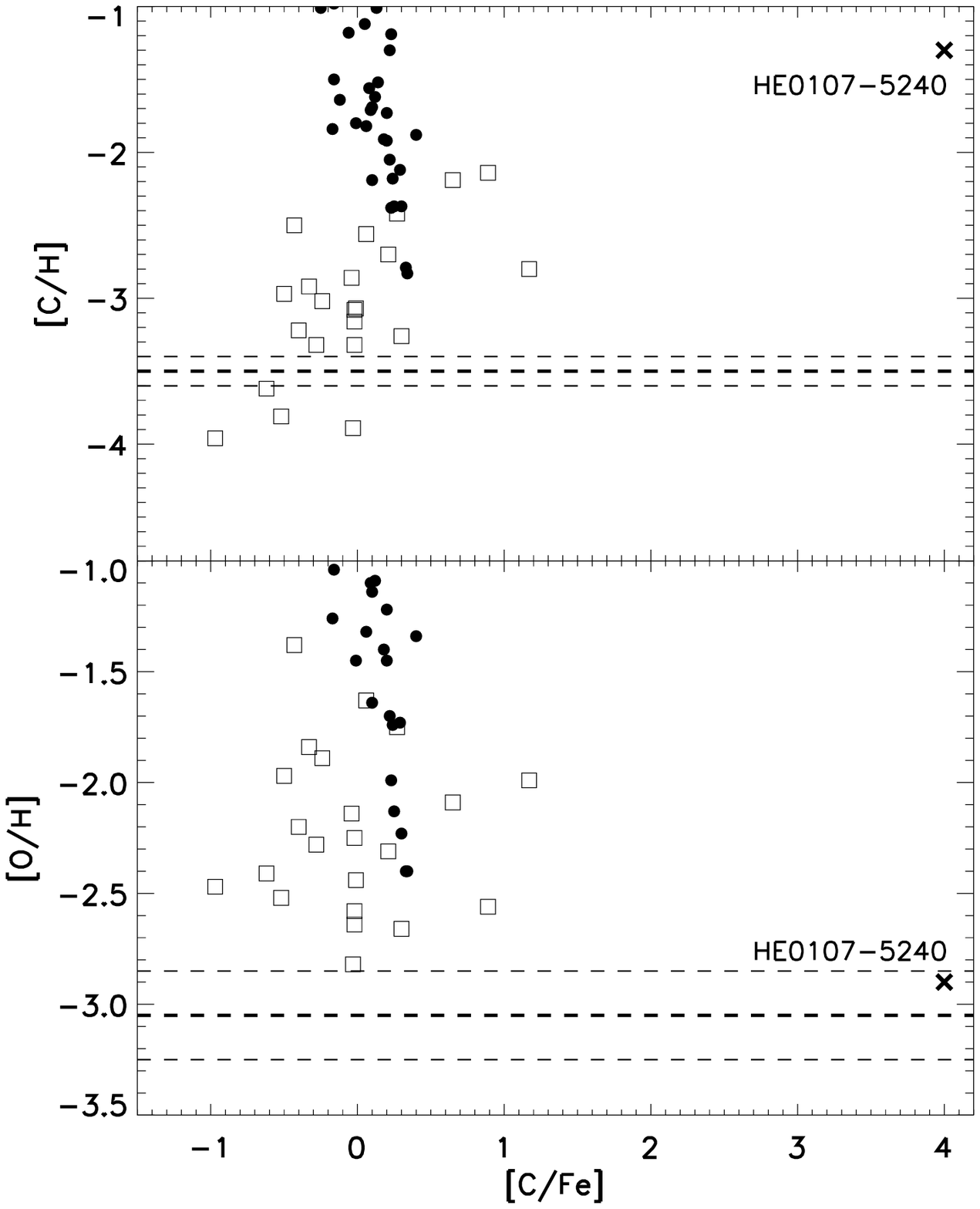}
\end{figure*}

\end{document}